\documentclass[12pt,preprint]{aastex}

\usepackage[usenames,dvips]{color}
\usepackage{epsfig}
\usepackage{psfig}
\usepackage{lscape}
\newcommand{\chisq}{$\chi^{2}$}
\newcommand{\psr}{PSR~J1723-2837}

\slugcomment{}

\shorttitle{High energy observation of MSP \psr}
\shortauthors{Hui et al.}

\begin{document}

\title{Exploring the X-ray and $\gamma-$ray properties of the redback millisecond pulsar \psr} 

\author{C. Y. Hui\altaffilmark{1}, P. H. T. Tam\altaffilmark{2}, J. Takata\altaffilmark{3}, A. K. H. Kong\altaffilmark{2}, K. S. Cheng\altaffilmark{3}, J. H. K. Wu\altaffilmark{2}, L. C. C. Lin\altaffilmark{4} and E. M. H. Wu\altaffilmark{3}}

\altaffiltext{1}{Department of Astronomy and Space Science, 
Chungnam National University, Daejeon 305-764, Korea}
\altaffiltext{2}
{Institute of Astronomy and Department of Physics,
National Tsing Hua University, Hsinchu, Taiwan }
\altaffiltext{3}
{Department of Physics, University of Hong Kong, Pokfulam Road,
Hong Kong}
\altaffiltext{4}{General Education Center, China Medical University, Taichung 40402, Taiwan}


\begin{abstract}
We have investigated the X-ray and $\gamma-$ray properties of the redback millisecond pulsar 
\psr\ with \emph{XMM-Newton}, \emph{Chandra} and \emph{Fermi}. We have discovered the X-ray orbital 
modulation of this binary system with the minimum that coincides with the phases of radio eclipse. 
The X-ray emission is clearly non-thermal in nature which can be well described by a simple 
power-law with a photon index of $\sim1.2$. The phase-averaged luminosity is $\sim9\times10^{31}$~erg/s 
in 0.3-10~keV which consumes $\sim0.2\%$ of the spin-down power. We have detected the $\gamma-$ray emission 
in $0.1-300$~GeV from this system at a significance of $\sim6\sigma$ for the first time. The $\gamma-$rays 
in this energy range consumes $\sim2\%$ of the spin-down power and can be modeled by a power-law with 
a photon index of $\sim2.6$. We discuss the high energy properties of the new redback in the context of 
an intrabinary shock model. 
\end{abstract}


\section{Introduction}
In the last few years, a new population of eclipsing binary millisecond pulsars (MSPs) has emerged. 
Although the range of orbital period spanned by these systems is similar to that of ``black widow" 
MSPs ($P_{b}\lesssim20$~hrs), their companion masses ($M_{c}\sim0.2-0.4M_{\odot}$) are significantly 
larger than that of black widows ($M_{c}\ll0.1M_{\odot}$) (Roberts 2013). This population is dubbed as 
``redbacks". Chen et al. (2013) suggest the determining factor for producing black widows or redbacks 
is the efficiency of companion evaporation. They further argue that redback systems do not evolve into 
black widows with time.

More than a dozen of redbacks have 
been identified so far.\footnote{see http://apatruno.wordpress.com/about/millisecond-pulsar-catalogue/ for updated information.} 
PSR~J1023+0038 is the first identified redback MSP which provides evidence for the transition from an X-ray binary to a radio MSP (Archibald et al. 2009, 2010; Thorstensen \& Armstrong 2005).
These systems are suggested to have their states possibly swinging between rotation and accretion power phases 
according to the mass transfer rate (Shvartsman 1970; Burderi et al. 2001). Recent observations of episodic accretion 
from the redbacks PSR~J1023+0038 (Patruno et al. 2013; Takata et al. 2013) and  
PSR~J1824-2452I (Papitto et al. 2013) are consistent with this scenario.

\psr, which is a redback, has its radio and optical properties been recently 
reported (Crawford et al. 2013). The dispersion measure suggests it locates at a 
distance of $0.75\pm0.10$~pc (Crawford et al. 2013).
Its rotational period and orbital period are 
1.86~ms and 14.8~hr, respectively. The spin-down luminosity of the pulsar is estimated as 
$\dot{E}=4.6\times10^{34}$~erg~s$^{-1}$. 
The mass of its companion lies in a range of $0.4-0.7M_{\odot}$ and optical spectroscopy 
indicates it is a G-type star. The pulsar follows a nearly circular orbit. The eclipse of 
radio pulses takes $\sim15\%$ of the orbit which is twice the Roche lobe size inferred for the
companion (Crawford et al. 2013). In this Letter, we report the results from our X-ray 
and $\gamma-$ray analysis of this system. 

\section{Observation \& Data Analysis} 
\subsection{{\it XMM-Newton} observations}
\psr\ has been observed by \emph{XMM-Newton} on 3 March 2011 for a total exposure of $\sim56$~ks 
with all EPIC cameras operated in full frame mode (Obs. ID: 0653830101) which provides a fairly 
uniform orbital coverage. With the updated 
instrumental calibrations, we generated all the event files with XMM Science Analysis 
Software (XMMSAS version 12.0.1). We selected only the data in the energy 
range of $0.3-10$~keV and those events correspond to the pattern in the range of $0-12$ for 
MOS1/2 cameras and $0-4$ for the PN camera. Bad pixels were also excluded. 
The effective exposures for MOS1, MOS2 and PN after filtering are found to be 
$\sim50.7$~ks, $\sim49.6$~ks and $\sim49.3$~ks respectively. 
Within a source region of 20~arcsec radius around the pulsar timing position 
(RA=17$^{\rm h}$23$^{\rm m}$23.1856$^{\rm s}$ Dec=-28$^{\circ}$37$^{'}$57.17$^{''}$ (J2000)), 
there are 4993~cts, 3655~cts and 13261~cts (source+background) extracted from 
MOS1, MOS2 and PN CCDs, respectively. All the EPIC data are found to be unaffected by CCD pile-up.

\begin{figure}[t]
\centerline{\psfig{figure=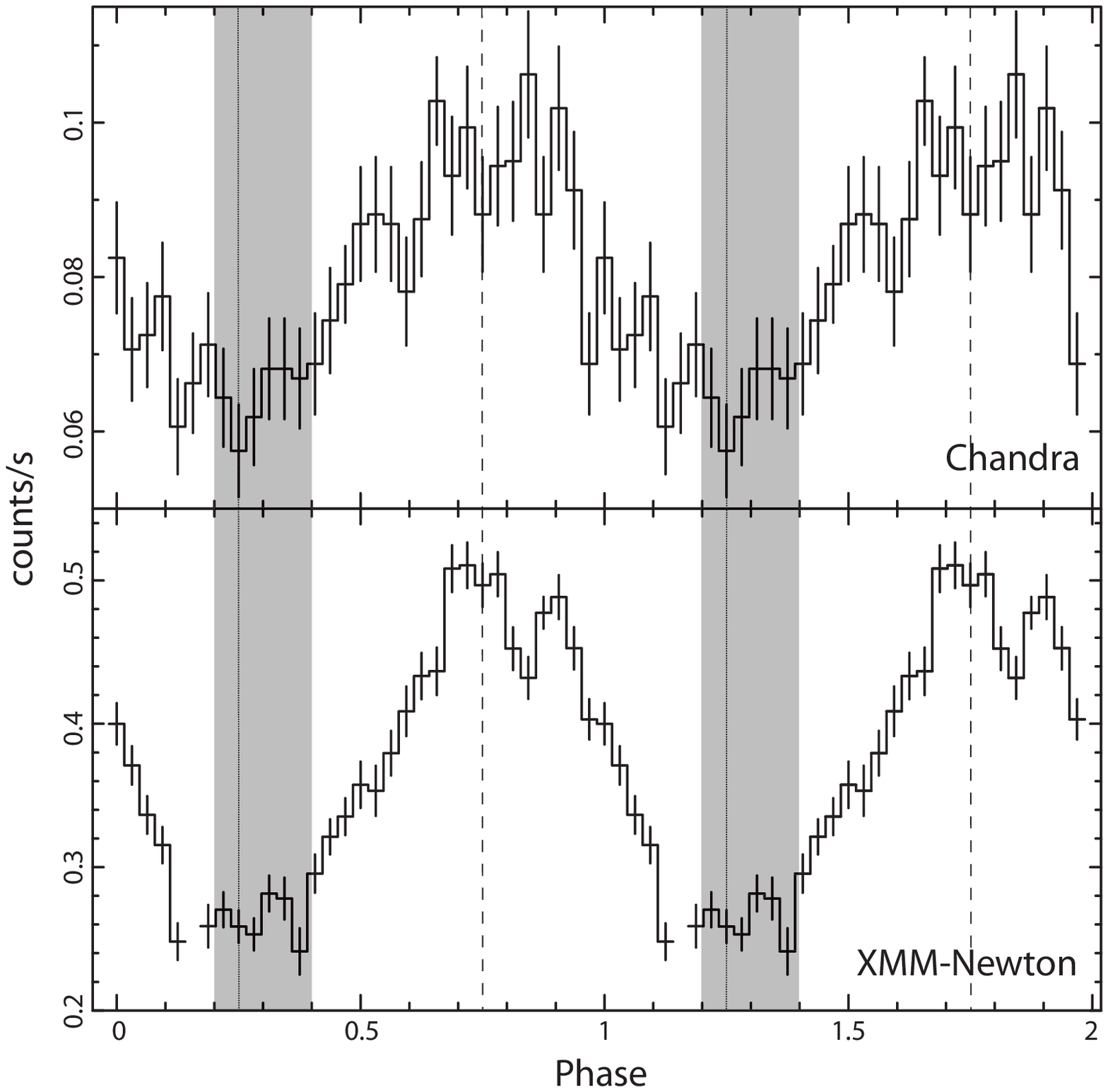,width=17cm,clip=}}
\caption[]{The background-subtracted light curves of \psr\ as observed 
by \emph{Chandra} ACIS in $0.3-7$~keV ({\it upper panel}) and 
by \emph{XMM-Newton} in $0.3-10$~keV with the data from all EPIC cameras combined ({\it lower panel}). 
The same data have been repeated for another orbital cycle for demonstrating the modulation clearly.  
The shaded region illustrates the range of the radio eclipse. The dotted line and the dashed line 
illustrate the phases of INFC and SUPC respectively.}
\label{lc}
\end{figure}

We firstly searched for the orbital modulation in X-ray. Before any temporal analysis performed, we have 
applied barycentric correction to the arrival times of all the events by using the 
updated planetary ephemeris JPL DE405. Then we subtracted the background in the individual 
camera by adopting an annular region with inner/outer radius of 25~arcsec/45~arcsec
centered at the pulsar position. Although the soft proton flares have contaminated the data 
in several regions of the CCDs, 
examining the light curve of the aformentioned small region-of-interest reveals no 
flare-like events. Therefore, we keep all data in order to optimize the orbital coverage 
for the analysis. 
Using the radio ephemeris for \psr\ (Tab.~2 in Crawford et al. 2013), we folded the background-subtracted 
light curves at 
the orbital period. The X-ray modulation can be clearly seen with all three cameras. In order 
to improve the photon statistics, we combine all the EPIC data and the resultant 
light curve is shown in Figure~\ref{lc}. We adopted the time of ascending node 
(MJD 55425.320466) to be the epoch of phase zero. Due to the incomplete orbital coverage after 
data filtering, there is gap at the orbital phase interval $0.14-0.18$ in this \emph{XMM-Newton} data. 

The minimum of the X-ray orbital modulation occurs at the same phase interval that radio eclipse 
has been observed in 1.4-3.1~GHz (cf. Fig.~1 in Crawford et al. 2013) which is illustrated by the shaded region 
in Figure~\ref{lc}. This interval encompasses the inferior 
conjunction (INFC) at a phase of $\sim0.25$ where the pulsar is behind its companion.
Apart from the region around INFC, some observations at 2~GHz peformed 
in the phase interval of $0.5-0.6$ 
also could not detect the pulsar (Crawford et al. 2013).
However, in examining the X-ray orbital modulation at this 
interval, we cannot identify any peculiar behavior. 
On the other hand, the maximum of the modulation is found in the orbital phase interval of $0.7-0.9$ 
which encompasses the superior conjunction (SUPC).
(see Fig.~\ref{lc}). In this interval, we have found a dip near a phase of $\sim0.8$. 

For investigating its X-ray spectral properties, we extracted the source and background spectra from the same
regions adopted in the temporal analysis. 
We grouped the spectra obtained individually from MOS1 and MOS2 so as to have at least 50 counts per bin. For the PN 
spectrum, we grouped it to have at least 60 counts per bin.
The spectra obtained from all three cameras are fitted 
simultaneously to the tested models. 
All the uncertainties quoted in this paper are $1\sigma$ for 2 parameters of interest (i.e. $\Delta$\chisq$=2.3$).

First, we have examined the phase-averaged X-ray spectrum of \psr. We found that a simple absorbed power-law 
model can describe the observed spectral data reasonably well (\chisq$=171.84$ for 179 d.o.f). 
The best-fit model yields a column density of $N_{H}=2.31^{+0.28}_{-0.25}\times10^{21}$~cm$^{-2}$, 
a photon index of $\Gamma_{X}=1.19\pm0.05$ and a 
normalization of $1.10^{+0.08}_{-0.07}\times10^{-4}$~photons~keV$^{-1}$~cm$^{-2}$~s$^{-1}$ at 1 keV. 
The best-fit $N_{H}$ is consistent with the 
value inferred from the optical extinction $A_{V}=1.2$ of this system 
(Crawford et al. 2013; Predehl \& Schmitt 1995).
The unabsorbed energy flux in $0.3-10$~keV is 
$f_{x}=1.32^{+0.19}_{-0.16}\times10^{-12}$~erg~cm$^{-2}$~s$^{-1}$. 
At a distance of 0.75~kpc, this corresponds to an X-ray luminosity of $L_{x}\sim8.9\times10^{31}$~erg~s$^{-1}$. 
We have also attempted to fit the X-ray spectrum of \psr\ with thermal models. 
While a black body yields a reduced \chisq\ much larger than unity, the thermal bremsstrahlung model results
in an unphysically high temperature ($kT>200$~keV).

\begin{figure}[t]
\centerline{\psfig{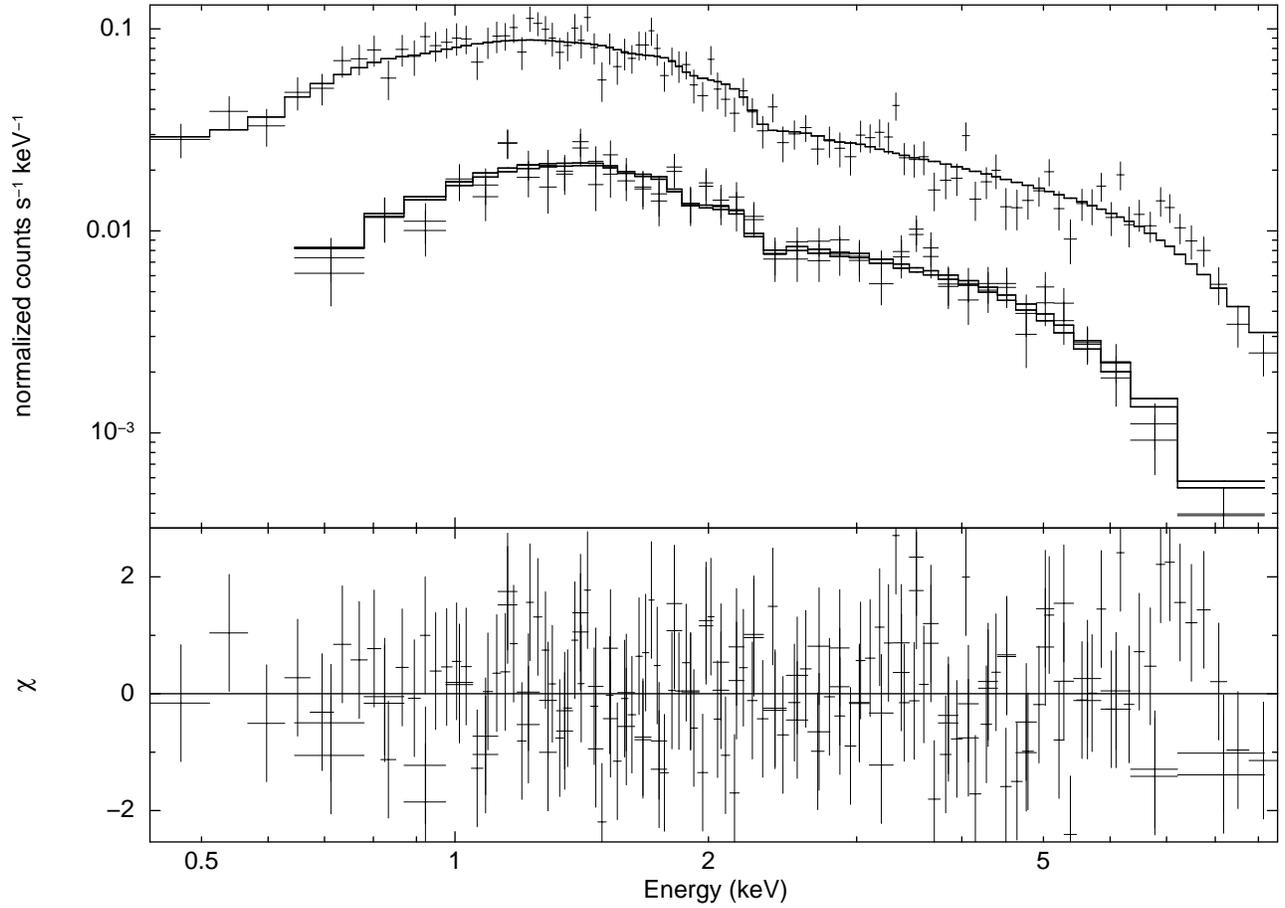}}
\caption[]{The phase-averaged X-ray spectra of \psr\ as observed by \emph{XMM-Newton} and 
simultaneously fitted to an absorbed power-law model (upper panel) and contribution to 
the \chisq-fit statistic (lower panel).}
\label{phase_ave}
\end{figure}

We proceeded to investigate if the spectral properties vary at different orbital phases. We have 
extracted the spectra from two phase ranges, 0.2-0.4 and 0.7-0.9 which encompass INFC and SUPC respectively.
For the INFC interval, the best-fit yields an $N_{H}=2.94^{+2.30}_{-1.25}\times10^{21}$~cm$^{-2}$ and 
$\Gamma_{X}=1.23^{+0.32}_{-0.19}$. On the other hand, the corresponding parameters for the SUPC interval  
are estimated as $N_{H}=2.01^{+0.49}_{-0.43}\times10^{21}$~cm$^{-2}$ and $\Gamma_{X}=1.09\pm0.09$. 
Within the statistical uncertainties, the spectral parameters inferred in these two intervals are 
consistent with the phase-averaged values. 
Based on this \emph{XMM-Newton} observation, there is no evidence of X-ray spectral variability across the orbit of 
\psr.

\subsection{{\it Chandra} observation}
{\it Chandra} observed \psr\ with the Advanced CCD Imaging Spectrometer (ACIS) on 2012 July 11 (ObsID: 13713). 
The orbital coverage is continuous with a sum of good time interval of 55ks, which is slightly
more than one orbital cycle.
The ACIS was operated with a sub-array mode to achieve a
timing resolution of 0.4s. 
Due to the loss of efficiency for the sub-array model, the effective exposure is shorter (49.9 ks).
Similar to the {\it XMM-Newton} observation, we performed barycentric correction 
to the arrival times of all the events by using the updated planetary ephemeris JPL DE405. \psr\ is clearly 
seen and is the brightest source in the image. In order to reduce the background, we limited our analysis in 
the 0.3--7 keV band. Using a circular source region with a radius of 4 arcsec, there are 4468 counts for subsequent 
analysis. An annular source free region centered at \psr\ was used for background subtraction.

We extracted the 0.3--7 keV lightcurve of \psr; similar to the {\it XMM-Newton} observation, the 14.8-hr orbital modulation 
is clearly seen (see Fig. 1). Although the number of counts is much less than that of {\it XMM-Newton}, the lightcurve 
profile is consistent with the {\it XMM-Newton}'s one. In spite of the large error bars, it is worth noting that there is a 
hint for the dip at phase $\sim0.8$ seen in the {\it XMM-Newton}.

For spectral analysis, we first investigated the phase-averaged spectrum. The spectrum can be well described 
(\chisq$=186.85$ for 204 d.o.f) with an absorbed power-law model with $N_H=(1.68\pm0.34)\times10^{21}$cm$^{-2}$ 
and $\Gamma_X=1.00\pm0.07$. The unabsorbed 0.3--10 keV flux is $1.64^{+0.04}_{-0.06}\times10^{-12}$ erg cm$^{-2}$ s$^{-1}$. 
The spectrum of {\it Chandra} is only slightly harder than that of {\it XMM-Newton}. Since \psr\ is bright for {\it Chandra}, 
some pile-up may contaminate the spectral analysis. We also included a pile-up model and found that the pile-up fraction 
is only 1.7\% and the spectral parameters are more or less the same. We then performed a phase-resolved spectroscopy using 
the same phase bins for INFC and SUPC as in the {\it XMM-Newton} data. Both spectra can be fit with an absorbed power-law. For INFC, 
the best fit is $N_H= (3.51^{+1.42}_{-1.15})\times10^{21}$cm$^{-2}$ and $\Gamma_X=1.35\pm0.22$, while for SUPC the corresponding 
parameters are $N_H= (1.3\pm1.0)\times10^{21}$cm$^{-2}$ and $\Gamma_X=0.82\pm0.17$. 
Like in the {\it XMM-Newton} data, there is no evidence for orbital dependent spectral variability. 

For better constraining the INFC and SUPC spectral shape, we simultaneously fitted the spectra obtained by 
{\it XMM-Newton} and {\it Chandra} in these two intervals. The cross-calibration was accounted by allowing the normalizations
to be different for these two telescopes. Assuming $N_{H}$ is variable at different orbital phases, we obtained 
$N_H=4.39^{+1.23}_{-1.00}\times10^{21}$cm$^{-2}$, $\Gamma_X=1.48^{+0.17}_{-0.14}$ 
and $N_H=2.01^{+0.47}_{-0.39}\times10^{21}$cm$^{-2}$, $\Gamma_X=1.05\pm0.08$ for INFC and SUPC respectively. 
The spectral parameters inferred in these phase intervals can be reconciled within $2\sigma$ uncertainties. 
We have also explored the possibility that there is no significant variation of $N_{H}$ across the orbit. We jointly 
fitted individual power-law model to the INFC and SUPC spectra with $N_{H}$ in the individual model tied together. This
yields $N_{H}=2.47^{+0.46}_{-0.42}\times10^{21}$cm$^{-2}$ and 
$\Gamma_X=1.23^{+0.11}_{-0.08}$, $\Gamma_X=1.12^{+0.08}_{-0.04}$ for INFC and SUPC respectively. The photon indices in these 
two intervals are consistent within $1\sigma$ uncertainties. Therefore, we do not found any conclusive evidence of the 
spectral variation across the orbit. 

\subsection{{\it Fermi} {\bf Large Area Telescope} observations}
The $\gamma$-ray data used in this work were obtained between 2008 August 4 and 2013 July 15, which are available at the Fermi Science Support Center~\footnote{\url{http://fermi.gsfc.nasa.gov/ssc/}}. We used the Fermi Science Tools v9r31p1 package to reduce and analyze the Pass7 data in the vicinity of \psr. To reduce the contamination from Earth albedo $\gamma$-rays, we excluded events with zenith angles greater than 100$^\circ$. We used events with energies between 100 MeV and 300 GeV in the binned likelihood analysis. The corresponding instrumental response functions were used. 

We first chose a rectangular region of dimension 28$^\circ\times$28$^\circ$ centered on the radio timing position as our region-of-interest (ROI). We subtracted the background contribution by including the Galactic diffuse model (gal\_2yearp7v6\_v0.fits) and the isotropic background (iso\_p7v6source.txt), as well as all sources in the second Fermi/LAT catalog (2FGL; Nolan et al. 2012) within the circular region of 25$^\circ$ radius around \psr\ in the background source model. We assumed the respective spectrum in the 2FGL catalog for the sources considered. The normalization of the diffuse components and spectral parameter values of sources within 5$^\circ$ from \psr~were allowed to vary. 

Using the ``source''-class events and assuming a power-law spectrum at the pulsar position, the maximized test-statistic (TS) value (Mattox et al. 1996) we obtained for the \psr~position is 38.3, corresponding to a detection significance of 6.2$\sigma$. The power-law index is $\Gamma_\gamma=2.6\pm0.1$. The photon flux from the power-law fit is $(3.5\pm0.7)\times10^{-8}$~photons~cm$^{-2}$~s$^{-1}$, and the energy flux is $(1.5\pm0.3)\times10^{-11}$~erg~cm$^{-2}$~s$^{-1}$, corresponding to the $\gamma$-ray luminosity of $L_{\gamma}\sim10^{33}$~erg~s~$^{-1}$ at a distance of 0.75~kpc. Since it is uncertain how much the pulsar proper motion contributes to the spin-down rate, the resulting spin-down luminosity can be treated as an upper limit of the intrinsic spin-down luminosity. Therefore, $L_{\gamma}$ would be no smaller than $\sim$2\% of the intrinsic spin-down luminosity. Attempts to fit the data using a power-law with an exponential cutoff at the \psr\ position did not provide a constraining parameter set. We also chose different ROIs to verify our results.

Given the proximity of \psr~to the Galactic Center where the Galactic $\gamma$-ray diffuse background is high, we also performed a binned likelihood analysis using ``clean''-class events in a smaller ROI of dimension 21$^\circ\times$21$^\circ$ centered on the radio timing position. The TS value thus obtained is 24.2, corresponding to a detection significance of 4.9$\sigma$. Taking into account the above likelihood results using ``source'' events and ``clean'' events as well as the $\gamma$-ray count map shown in Fig. 3, we believe that the detection of the $\gamma$-ray source is robust. 

\begin{figure}[t]
\centerline{\psfig{figure=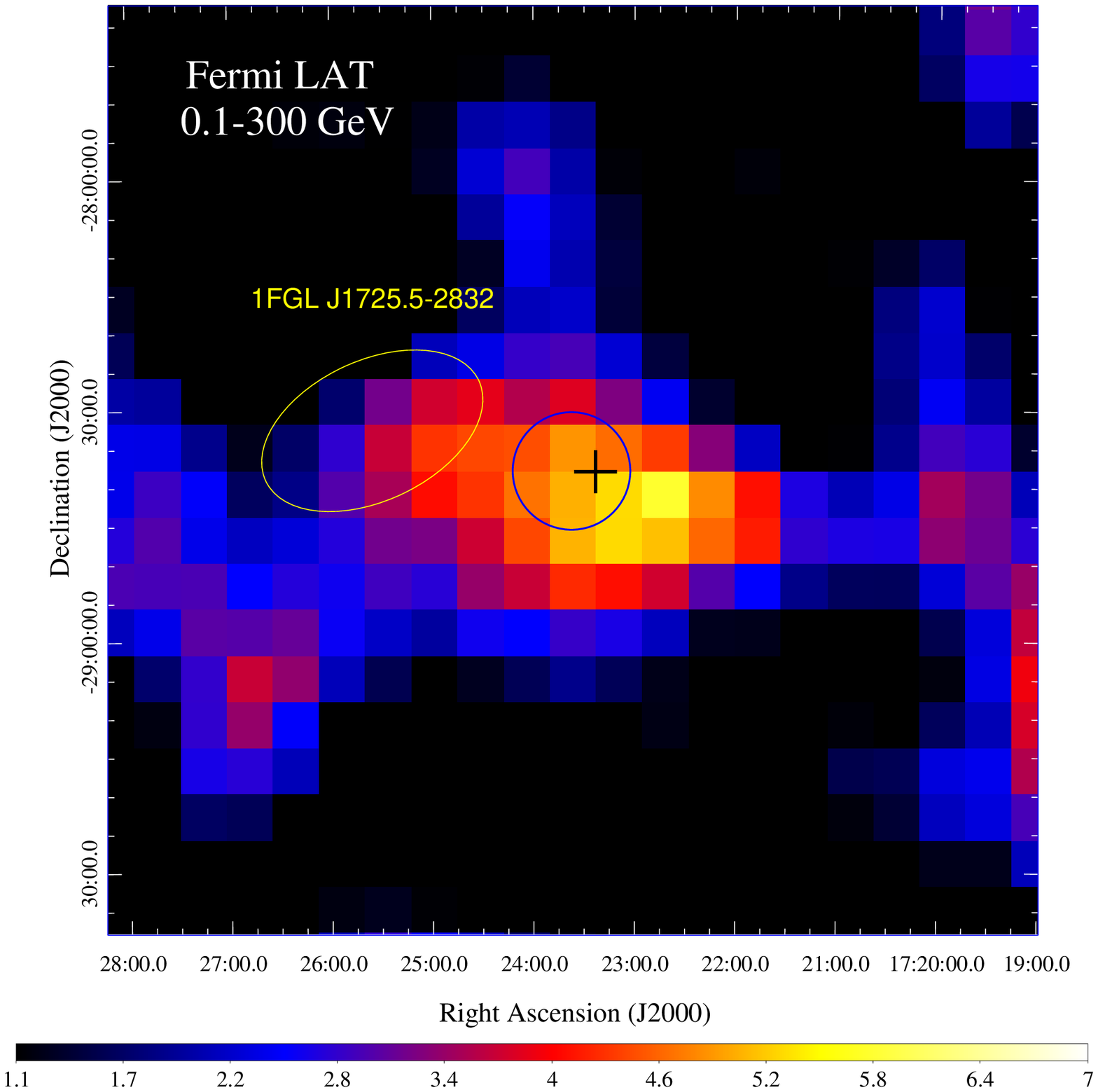,width=18cm,angle=0,clip=}}
\caption[]{The background-subtracted 0.1--300~GeV $\gamma$-ray count map, smoothed with a Gaussian width of 0$\fdg$3, of the 2$^\circ\times$2$^\circ$ region centered on \psr, whose radio timing position is indicated by the black cross. The blue circle indicates the error circle of the best-fit postion at the 68\% confidence-level. The error eclipse of 1FGL~J1725.5-2832, which is not regarded as a background source (see Sect. 2.3), is shown.}
\label{gamma_cmap}
\end{figure}

To better estimate the $\gamma$-ray position, we performed an unbinned analysis of the front-converted ``clean''-class events larger than 500~MeV using {\it gtfindsrc}, resulting in the position at RA(J2000)=260$\fdg$906, Dec(J2000)=$-$28$\fdg$631, with statistical uncertainty 0.13$^\circ$ at the 68\% confidence level, which is consistent with the position of \psr. 
The background-subtracted count map of the 2$^\circ\times$2$^\circ$ region centered on the radio position, constructed using the ``clean'' class events, is shown in Figure~\ref{gamma_cmap}.

A nearby source in the first Fermi catalog (Abdo et al., 2010), 1FGL~J1725.5-2832, is only $\sim$1$^\circ$ away from \psr\ (see Fig.~\ref{gamma_cmap}). We studied their relation by adding 1FGL~J1725.5-2832 to the background source model that had been used in the aforementioned analysis. It turned out that essentially the same group of photons contribute to the TS values of both sources. Therefore, we treat 1FGL~J1725.5-2832 to be the same $\gamma$-ray source associated with \psr.

Using the radio ephemeris reported in Crawford et al. (2013), we searched for possible $\gamma$-ray variability related to the orbital modulation. We divided the photons into two phase-intervals, one centered at the SUPC (0.5-1.0), the other centered at the INFC (0.0-0.5), and performed binned likelihood analysis for each group of photons, using ``source'' events and same analysis cuts and source model as described in the full-orbit analysis. No difference of the spectral parameters were found between SUPC and INFC. We also searched for any periodic signals using $H$-test (de Jager \& B\"usching 2010). Neither the $\gamma$-ray periodicity related to the orbital period nor the rotation period was found. However, the bright Galactic diffuse emission, uncertainties in the ephemeris used over the Fermi data span, and small photon statistics from the source make our periodicity searches less sensitive.

\section{Discussion}
With a simple one-dimensional model (Takata et al. 2012), we constrain 
the pulsar wind parameters of \psr. 
Since the radio eclipse at several GHz lasts 
only ~15\% of its orbit (Crawford et al. 2013), we may expect 
the shock wraps the companion star and its distance from the pulsar 
is on the order of the semi-major axis, 
which is $a\sim  3.7\times 10^{10}/\sin i$ ~cm, where $i$ is 
the orbital inclination angle. Based on the measurement of the Doppler velocity of the companion, Crawford et al. (2013) 
suggested the inclination angle of  $i\sim 30^{\circ}-41^{\circ}$ for \psr. 
The radio eclipse also suggests the angle of  flow from the companion star 
measured from the pulsar is $\theta_a\sim 2\pi\times 0.15=0.94$. 
If the pulsar wind is
 emitted spherically then the fraction of the pulsar wind stopped 
by the outflow is $\sim 5\%$.  
The magnetic fields upstream  ($B_1$) and behind ($B_2)$ 
 the shock are estimated as $B_1\sim (L_{sd}\sigma/a^2c)^{1/2}$ 
and $B_2=3B_1$, where $\sigma$ is the ratio of the magnetic energy to kinetic 
energy of the cold relativistic pulsar wind and we assume it to be 
smaller than unity (i.e. kinetically dominated flow). At the shock,
 the pulsar wind particles are accelerated beyond the Lorentz factor 
($\Gamma_1$) of the cold relativistic pulsar wind in the upstream region. 
For the maximum Lorentz factor of the accelerated particles, it is determined by 
balancing the accelerating time scale, $\tau_{acc}\sim \Gamma m_ec/(eB_2)$, 
and the synchrotron cooling time scale, $\tau_s\sim 9m_e^3c^5/(4e^4B^2\Gamma)$.

The observed X-ray photon index $\Gamma_{X}\sim 1.2$ may imply 
the power law index of the particle distribution at
 the shock is  $p\sim 1.4$. For 
the usual power law index  $p\sim 2-2.5$ of the shock acceleration, 
this hard photon index in the spectrum 
 will be reproduced if  the typical synchrotron 
energy ($E_1$) of the minimum Lorentz factor $\Gamma_1$ is larger than 10~keV. 
The photons index changes from $\alpha =(p+1)/2$ above $E_1$ to 
$\alpha=3/2$ below $E_1$. In such a case, the X-ray emission properties 
break the degeneracy of $\sigma$ and $\Gamma_1$ in the emission model. For example, we can fit the observed properties $\alpha\sim 1.2$ and $f_{x}\sim 
1.3\times 10^{-12}{\rm erg~cm^{-2}~s^{-1}}$ with 
($\sigma,~\Gamma_1)\sim (0.1,~5\times 10^4)$. The predicted Lorentz 
factor $\Gamma_1\sim 5\times 10^4$ is similar to those of 
the original black widow pulsar PSR B1957+20 (Wu et al. 2012). 
As there is no evidence of spectral break found in 0.3-10~keV, observations 
with state-of-the-art hard X-ray telescopes, {\it NuSTAR} and the upcoming Astro-H 
will be important for constraining $E_{1}$. 

Because PSR J1723-2837 follows an almost circular orbit, the shock distance 
from the pulsar does not vary across the orbit, suggesting the spectral properties 
of the intrinsic shock  emissions do not modulate with
 the orbital phase. The variation of the observed flux will be caused 
by either Doppler boosting effect with a mildly relativistic flow of the 
shocked flow or physical eclipse of the emission region. However, these
two effects will not produce a significant variation in the spectrum. 
The observed amplitude (see Figure~1) implies the Doppler factor is $\sim 1-2$, 
which does not cause a significant change in the spectral shape. 
Furthermore, the synchrotron cooling time scale of the particles that
emit photons of energies $<10$~keV 
is $\tau_{s}\sim 33(B_2/10 {\rm G})^{-3/2}(E_{s}/10{\rm keV})^{-1/2}$s, 
which is longer 
than the crossing time scale of the emission region, $\tau_c\sim 1$s.  
The distribution of the particles and therefore the spectral shape of X-ray 
emission in $<10~keV$  does not evolve in the emission region.
As a result, even if the partial region of the emission
region is covered by the star, the observed spectral shape is the same with 
that for whole emission regions. 

For black widow/redback pulsars, the magnetospheric emissions and pulsar wind
 emissions produce the GeV gamma-rays. For the original black widow pulsar PSR
 B1957+20, Wu et al. (2012) suggested both the magnetospheric and the pulsar 
wind emissions contribute to the GeV emissions seen by {\it Fermi}. 
With the results described in section~2.3,  it is unclear which process 
operates for the observed gamma-rays for PSR~J1723-2837, because neither 
pulsar's spin period nor orbital modulation was founded in 
the current {\it Fermi} data, and whether or not there is a cutoff in the spectrum 
is inconclusive. A single power-law feature in 
the observed spectrum in 0.1--300~GeV band
would suggest that the gamma-rays are emitted by the  pulsar wind
 particles  accelerated at the shock. In this case, the synchrotron 
radiation process explains the observed emissions around 100~MeV and 
the inverse-Compton process off the stellar photons contributes to  
the emissions above $\sim 10$~GeV. 

The shocked particles that emit the synchrotron photons will produce 
TeV photons through the inverse-Compton off the stellar photons, 
for which the effective temperature is $T_{eff}\sim 4800-6000$K.  Based on 
the calculation with the isotropic photon fields, we can estimate the 
flux above $>100$GeV as $F_{>100GeV}\sim 5\times 10^{-13}{\rm erg~cm^{-2}~s^{-1}}$, 
which could possibly be measured by 
the planned Cherenkov Telescpe Array (CTA). The measurement of the CTA on 
the redback systems will provide us additional information such as 
Lorentz factor of the relativistic pulsar wind and the maximum Lorentz factor of the shocked particles, etc.

\acknowledgments{
CYH is supported by the National Research Foundation of Korea through grant 2011-0023383. PHT is supported by the National Science Council of the Republic of China (Taiwan) through grant NSC101-2112-M-007-022-MY3.
JT and KSC are supported by a GRF grant of HK Government under HKU7009
11P. 
AKHK is supported by the National Science Council of the
Republic of China (Taiwan) through grant NSC100-2628-M-007-002-MY3 and
NSC100-2923-M-007-001-MY3. 
LCCL is supported by the National Science Council through grants NSC 101-2112-M-039-001-MY3.
}


\end{document}